\documentclass[12pt]{article}

\usepackage{color}
\usepackage{cite}
\usepackage{amsmath}
\usepackage{graphicx}

\usepackage{here}

\usepackage{soul}

\usepackage{ifpdf}
\ifpdf
  \DeclareGraphicsExtensions{.pdf,.png,.jpg}
\else
  \DeclareGraphicsExtensions{.ps,.eps}
\fi

\def\css{\hbox{$c\kern0.03em(\kern-0.1em ss\kern-0.1em)$}}
\def\bss{\hbox{$b\kern0.03em(\kern-0.1em ss\kern-0.1em)$}}
\def\qss{\hbox{$q\kern0.03em(\kern-0.1em ss\kern-0.1em)$}}
\def \barM{M\kern-0.84em\raise0.8em\hbox{\vrule width 9.5pt height 0.6pt}}

\def \beq{\begin{equation}}
\def \eeq{\end{equation}}
\def\eqref#1{(\ref{#1})}
\def\bea{\begin{eqnarray}}
\def\eea{\end{eqnarray}}

\def\jpsi{J\kern-0.15em/\kern-0.15em\psi\kern0.15em}

\def\Sd{S_{\kern-0.1em ss}}
\def\SSd{{\bold S}_{\kern-0.1em ss}}
\def\Sdi{S^i_{\kern-0.1em ss}}

\def\URLtilde{\lower0.2em\hbox{$\tilde{\phantom{a}}$}}
\def\mycomm#1{\hfill\break\strut\kern-3em{\color{red}\tt ====> #1
\color{black}}\hfill\break}

\newcount\timecount
\newcount\hours \newcount\minutes  \newcount\temp \newcount\pmhours
\hours = \time
\divide\hours by 60
\temp = \hours
\multiply\temp by 60
\minutes = \time
\advance\minutes by -\temp
\def\hour{\the\hours}
\def\minute{\ifnum\minutes<10 0\the\minutes
\else\the\minutes\fi}
\def\clock{
\ifnum\hours=0 12:\minute\ AM
\else\ifnum\hours<12 \hour:\minute\ AM
\else\ifnum\hours=12 12:\minute\ PM
\else\ifnum\hours>12
\pmhours=\hours
\advance\pmhours by -12
\the\pmhours:\minute\ PM
\fi
\fi
\fi
\fi
}

\def\monthname{\relax\ifcase\month 0/\or January\or February\or
March\or April\or May\or June\or July\or August\or September\or
October\or November\or December\else\number\month/\fi}
\def\today{\monthname~\number\day, \number\year}

\def\bold#1{\boldsymbol{#1}}

\def\draft{\color{red}
$\bold{\strut\kern-3em
\hbox{\tt \Large DRAFT, NOT TO BE DISTRIBUTED:  \clock, \today.}
}$\par\noindent\color{black}}
\begin{document}
\setcounter{footnote}{1}
\rightline{EFI 20-9}
\rightline{arXiv:2005.12424}
\vskip1.5cm
\centerline{\large \bf \boldmath INTERPRETATION OF EXCITED $\Omega_b$ SIGNALS
\unboldmath}
\bigskip

\centerline{Marek Karliner$^a$\footnote{{\tt marek@tauex.tau.ac.il}},
and Jonathan L. Rosner$^b$\footnote{{\tt rosner@hep.uchicago.edu}}}
\medskip

\centerline{$^a$ {\it School of Physics and Astronomy}}
\centerline{\it Raymond and Beverly Sackler Faculty of Exact Sciences}
\centerline{\it Tel Aviv University, Tel Aviv 69978, Israel}
\medskip

\centerline{$^b$ {\it Enrico Fermi Institute and Department of Physics}}
\centerline{\it University of Chicago, 5640 S. Ellis Avenue, Chicago, IL
60637, USA}
\bigskip
\strut

\begin{center}
ABSTRACT
\end{center}
\begin{quote}
Recently LHCb reported the discovery of four extremely narrow excited
$\Omega_b$ baryons decaying into $\Xi_b^0 K^-$.  We interpret these baryons as
bound states of a $b$-quark and a $P$-wave $ss$-diquark. For such a system
there are exactly five possible combinations of spin and orbital angular
momentum.  We predict two of spin 1/2, two of spin 3/2, and one of spin 5/2,
all with negative parity.  We favor identifying the observed states as
those with spins 1/2 and 3/2, and give a range of predicted masses for the one
with spin 5/2.  We update earlier predictions for these states based on the
five narrow excited $\Omega_c$ states reported by LHCb.  An alternative picture
of the states in which one of $J=1/2$ is extremely wide and hence not seen by
LHCb is discussed.
\end{quote}
\smallskip

\leftline{PACS codes: 12.39.Jh, 13.30.Eg, 14.20.Mr}


\section{Introduction \label{sec:intro}}

Recently LHCb reported the discovery of four extremely narrow excited
$\Omega_b$ baryons decaying into $\Xi_b^0 K^-$ \cite{Aaij:2020cex},
with masses and widths shown in Table \ref{tab:omb}.  We quote also our
favored spin-parity assignment for these states.  This result
follows upon the earlier observation by LHCb of five very narrow excited
$\Omega_c$ baryons \cite{Aaij:2017nav}, which we interpreted as $P$-wave
excitations of the $ss$ diquark with respect to the $c$ quark
\cite{Karliner:2017kfm}.
The global significance of the two lowest states is below $3 \sigma$, so our
assignments are subject to possible change with additional data.

\begin{table}
\caption{Masses, widths, and 90\% (95\%) confidence level upper limits on
natural widths of $\Omega_b = bss$ candidates reported by the
LHCb Collaboration \cite{Aaij:2020cex}.  The proposed values of spin-parity
$J^P$ are ours.
\label{tab:omb}}
\begin{center}
\begin{tabular}{c c c c c c} \hline \hline
State & Mass (MeV) & Width (MeV) & Proposed $J^P$
 &\multicolumn{2}{c}{Significances ($\sigma$)}  \\
 & & & & Local & Global \\ \hline
\vrule width 0pt height 2.5ex 
$\Omega_b(6316)^0$ & $6315.64 \pm 0.31 \pm 0.07 \pm 0.50$ &
 $<2.8 (4.2)$ & $1/2^-$ & 3.6 & 2.1 \\ 
$\Omega_b(6330)^0$ & $6330.30 \pm 0.28 \pm 0.07 \pm 0.50$ &
 $<3.1 (4.7)$ & $1/2^-$ & 3.7 & 2.6 \\
$\Omega_b(6340)^0$ & $6339.71 \pm 0.26 \pm 0.05 \pm 0.50$ &
 $<1.5 (1.8)$ & $3/2^-$ & 7.2 & 6.7 \\
$\Omega_b(6350)^0$ & $6349.88 \pm 0.35 \pm 0.05 \pm 0.50$ &
 $<2.8 (3.2)$ & $3/2^-$ & 7.0 & 6.2 \\
\vrule width 0pt depth 1.1ex 
 & & $1.4^{+1.0}_{-0.8} \pm 0.1$ & \\ \hline \hline
\end{tabular}
\end{center}
\end{table}
The discovery of the five excited $\Omega_c$ states raised some questions,
which we addressed: 

\begin{itemize}

\item[(a)] {\it Why five states?  Are there more in the $css$ system?}  There
are exactly five $1P$ excitations if the $ss$ diquark remains in its
color-triplet spin-1 ground state.  In an alternative picture the three
lowest states are $1P$ excitations while the two highest are $1/2^+$ and
$3/2^+$ $2S$ radial excitations.

\item[(b)] {\it Why are they so narrow?}  States with no nonstrange quarks
($u$ or $d$) do not couple directly to pions, closing important low-threshold
channels.

\item[(c)] {\it What are their spin-parity assignments?} We favored $J^P =
(1/2^-,1/2^-, 3/2^-,3/2^-,\\ 5/2^-)$ for the observed states, in order of
increasing mass.  An alternative assignment was $(3/2^-,3/2^-,5/2^-,1/2^+,
3/2^+)$.

\item[(d)] {\it Can one understand the mass pattern?} Yes; the favored pattern,
based on contributions of spin-orbit, spin-spin, and tensor force interactions,
was uniquely selected out of 5! = 120 possible permutations of the five
states.

\item[(e)] {\it Are there other similar states with different quark content,
in particular very narrow excited $\Omega_b$ baryons?}  LHCb has now observed
four out of the five predicted $1P$ excitations \cite{Aaij:2020cex}, leaving
a fifth to be predicted and observed.

\end{itemize}
The same questions can be asked for the four observed $\Omega_b$ states.  Which
of the expected five $\Omega_b$ states is missing, and what is its mass?  Is
the spin-weighted average of the $1P$ excitations consistent with expectation?

In Sec.\ \ref{sec:PW} we comment on $P$-wave $bss$ baryons.  We then
analyze spin-dependent forces for the $bss$ system in Sec.\ \ref{sec:sd},
building upon similar results \cite{Karliner:2017kfm} obtained previously
for the negative-parity $\Omega_c$ states.  We evaluate the energy cost for a
$P$-wave $bss$ excitation in Sec.\ \ref{sec:SP}, compare our present results
with our earlier predictions for the $\Omega_b$ system in Sec.\ \ref{sec:omb},
discuss alternative interpretations of the spectrum in Sec.\ \ref{sec:alt},
and conclude in Sec.\ \ref{sec:concl}.

\section{\boldmath $P$-wave \bss\ system \unboldmath \label{sec:PW}}

We retrace steps in \cite{Karliner:2017kfm} leading to five excitations of
the $ss$ diquark in a relative P wave with respect to a $b$ quark.  Consider
the $(ss)$ in \bss\ to be an $S$-wave color ${\bf\bar 3}_c$ diquark.  Then it
must have spin $\Sd = 1$.  This spin can be combined with the spin 1/2 of the
$b$ quark to a total spin $S=1/2$ or 3/2.  States with relative orbital
angular momentum $L=1$ between the spin-1 diquark and the $b$ quark are:
\bea
(L=1) \otimes (S=1/2) &=&(J=1/2,3/2)~, \nonumber \\
(L=1) \otimes (S=3/2) &=& (J=1/2,3/2,5/2)~.
\eea
All five states have negative parity $P$.  Those with $J^P=1/2^-$ decay to
$\Xi_b^0 K^-$ in an $S$-wave, while those with $J^P = 3/2^-,5/2^-$ decay to
$\Xi_b^0 K^-$ in a $D$-wave.

The LHCb experiment sees only four of the predicted five $P$-wave excitations
in the $\Omega_b = bss$ system \cite{Aaij:2020cex}.  Only four of the five
predicted $\Omega_c$ states are seen by Belle in $e^+ e^-$ collisions
\cite{Yelton:2017qxg}; the omitted state is the heaviest, $\Omega_c(3119)$.
This makes sense as kinematic suppression is greatest for the heaviest state.
For an initial state with no heavy flavor, the minimum mass recoiling against
a $css$ state such as $\Omega_c(3119)$ is $M(\Omega_c) = 2695.2 \pm 1.7$ MeV
while typical $e^+e^-$ c.m.s.\ energy is $M(\Upsilon(4S)) = 10579.4 \pm
1.2$ MeV \cite{Tanabashi:2019}.  In keeping with our identification of the
$\Omega_c(3119)$ as the state with $J^P = 5/2$, we shall assume that it is the
$J^P = 5/2^-~\Omega_b$ which is missing, and focus on the mass range above
$M(\Omega_b(6350))$ for it.

\section{Spin-dependence of masses \label{sec:sd}}

The masses of the $P$-wave excitations of the $ss$ diquark with respect to $b$
are split by spin-orbit forces, a tensor force, and hyperfine interactions,
leading to a spin-dependent potential \cite{Karliner:2017kfm}
\beq \label{eqn:vsd}
V_{SD} = a_1{\bold L} \cdot \SSd + a_2{\bold L} \cdot {\bold S_Q}
 + b [ - \SSd \cdot {\bold S_Q}
 + 3(\SSd \cdot {\bold r})({\bold S_Q} \cdot {\bold r})/r^2]
 + c \SSd \cdot {\bold S_Q}~.
\eeq

States with the same $J$ but different $S$ mix with one another, so the
mass shift operators $\Delta {\cal M}_{1/2,3/2}$ may be written as $2 \times 2$
matrices in bases labeled by $S = 1/2,3/2$:

\beq \label{eqn:m12}
\Delta {\cal M}_{1/2} = \left[ \begin{array}{c c} \frac13 a_2 - \frac43 a_1 &
\frac{\sqrt{2}}{3} (a_2-a_1) \\ \frac{\sqrt{2}}{3}(a_2-a_1) &
 - \frac53 a_1 - \frac56 a_2
\end{array} \right] +b \left[ \begin{array}{c c} 0 & \frac{1}{\sqrt{2}} \\
\frac{1}{\sqrt{2}}& -1 \end{array} \right] + c \left[ \begin{array}{c c} -1 &
 0 \\ 0 & \frac12  \end{array} \right]~,
\eeq
\beq \label{eqn:m32}
\Delta {\cal M}_{3/2} = \left[ \begin{array}{c c} \frac23 a_1 - \frac16 a_2 &
\frac{\sqrt{5}}{3}(a_2-a_1) \\ \frac{\sqrt{5}}{3}(a_2-a_1) &
 - \frac23 a_1 - \frac13 a_2
\end{array} \right] +b \left[ \begin{array}{c c} 0 & -\sqrt{5}/10 \\
 -\sqrt{5}/10 & \frac45 \end{array} \right] + c \left[ \begin{array}{c c} -1 &
 0 \\ 0 & \frac12 \end{array} \right]~,
\eeq
\beq \label{eqn:m52}
\Delta {\cal M}_{5/2} = a_1 + \frac12 a_2 - \frac15 b + \frac12 c~.
\eeq
The spin-weighted sum of these mass shifts is zero:
\beq \label{eqn:wtsum}
\sum_{J} (2J+1) \Delta {\cal M}_J = 0~.
\eeq
Note that the sums of eigenvalues of $\Delta {\cal M}_{1/2}$ and $\Delta {\cal
M}_{3/2}$ are equal to the traces of the corresponding matrices, making the
verification of Eq.\ (\ref{eqn:wtsum}) simple.

There are four measured masses and four independent parameters leading
to four mass {\it shifts} with respect to a spin-weighted average for which
one needs the fifth mass.  Thus the determination of the constants $a_1,a_2,%
b,c$ has one free parameter which we may take as ${\cal M}_{5/2}$.  We identify
the four known masses as shown in Table \ref{tab:omb}.

The spin-weighted average mass 
$\barM = \sum_J [(2J+1) {\cal M}_J]/18$
is linear in the unknown mass ${\cal M}_{5/2}$, with slope 1/3.
Anticipating the optimal fit ${\cal M}_{5/2} = 6358$ MeV (cf. the discussion
following Eq.~\eqref{eqn:pred}), $\barM$ can be rewritten in terms of the
deviation from this fit,
\beq
 \barM = 6344~{\rm MeV} + (1/3)({\cal M}_{5/2} - 6358~{\rm MeV})~.
\eeq
The limited range of $\barM$
will be of use when we study the $P$-wave excitation energy.

We now determine the parameters $a_1, a_2, b, c$ from the masses in Table
\ref{tab:omb}.  The measured masses permit one to write two identities
which are helpful in finding solutions.  We denote the two eigenvalues
of ${\cal M}_{1/2}$ by $M_1$ and $M_2$, and the two eigenvalues of
${\cal M}_{3/2}$ by $M_3$ and $M_4$.  A shorthand for ${\cal M}_{5/2}$ is $M_5$.
We find
\bea
a_2 & = & -\frac{4}{9}(M_1+M_2) -\frac{2}{9}(M_3+M_4) + \frac{4}{3}M_5
  - c - \frac{8}{3}a_1 \\
b & = & -\frac{5}{3}a_1 - \frac{5}{9}(M_1+M_2) + \frac{5}{9}(M_3+M_4)~.
\eea

\begin{figure}
 \includegraphics[width = 0.49\textwidth]{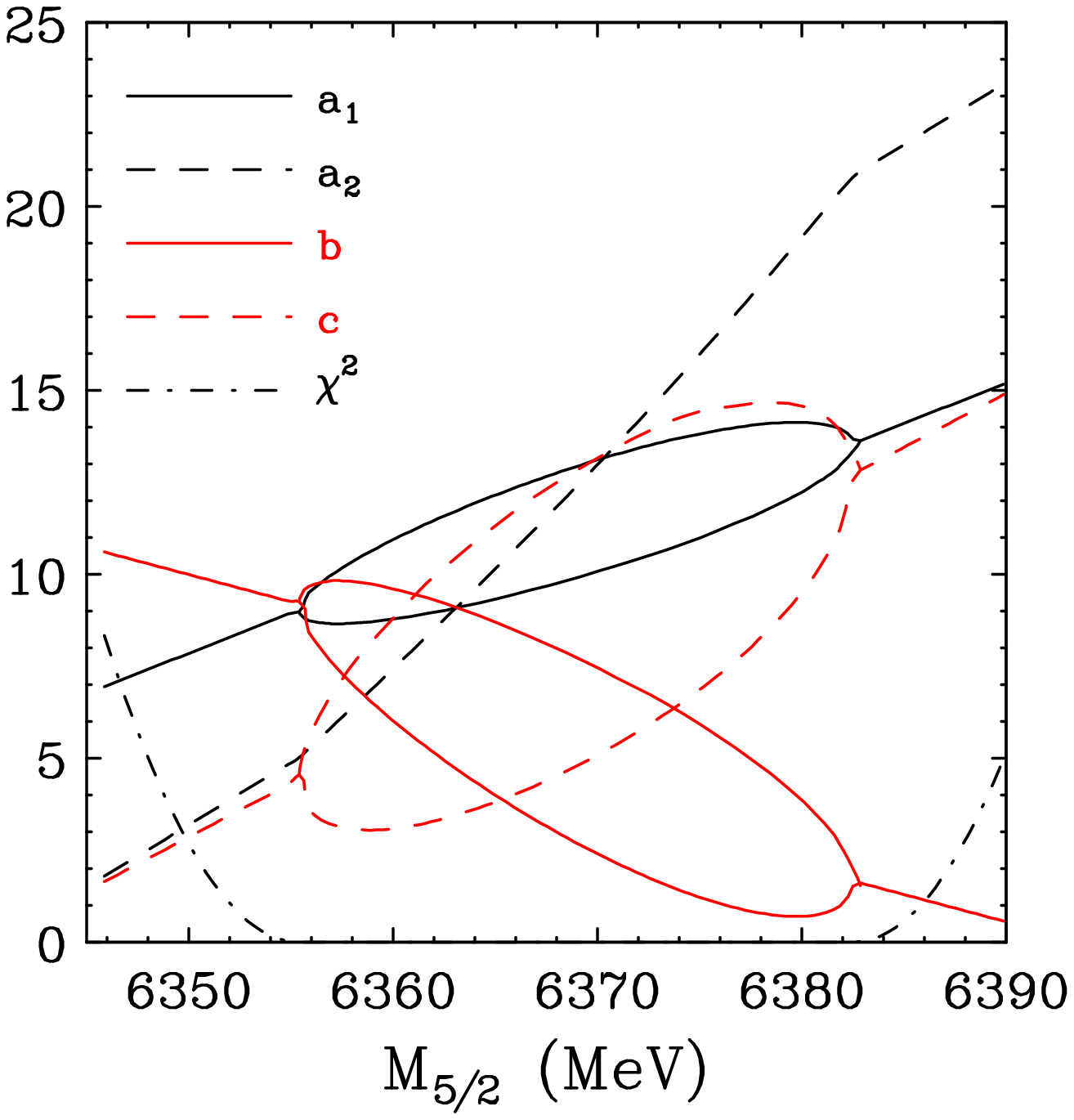}
 \includegraphics[width = 0.49\textwidth]{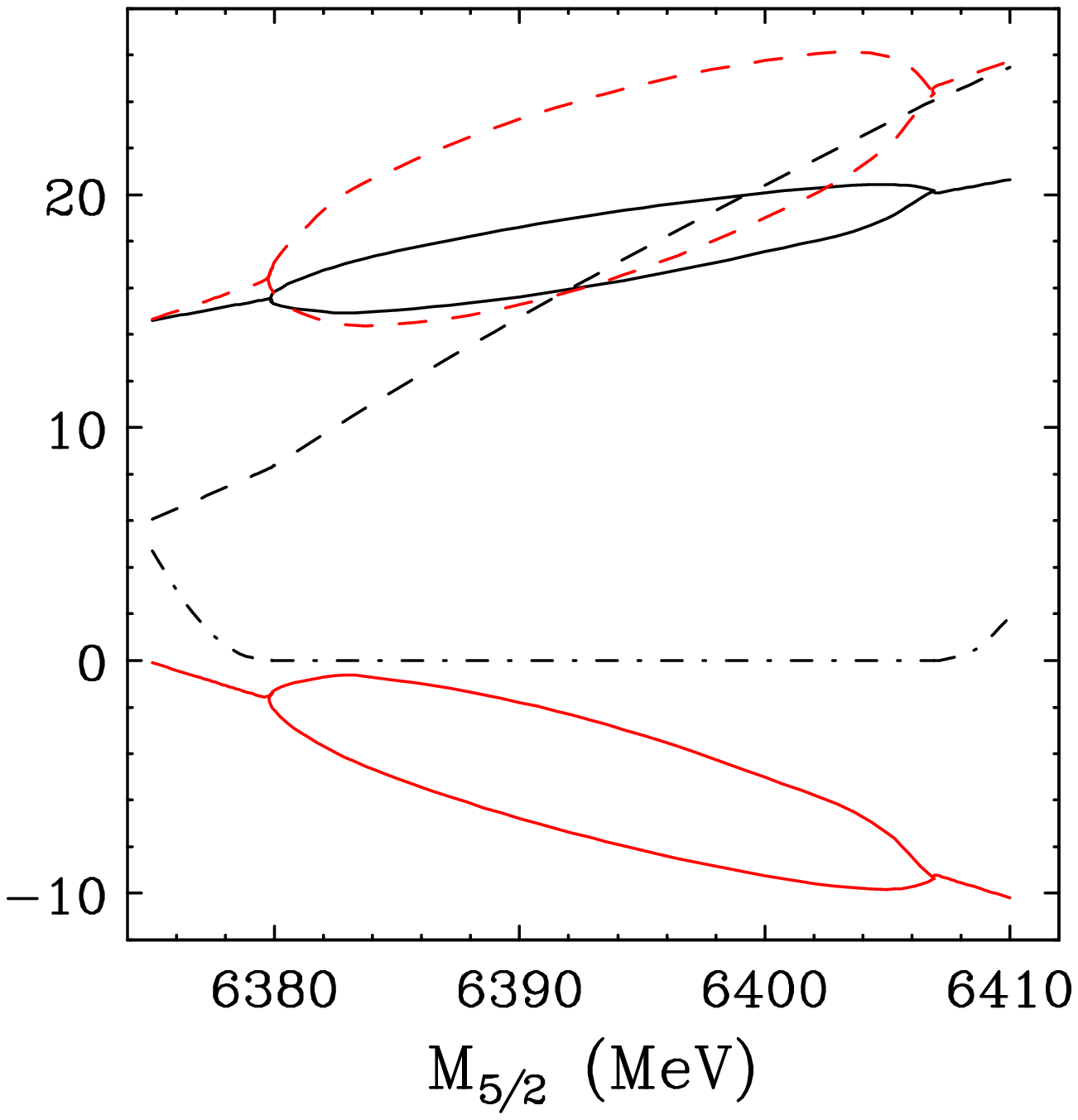}
\caption{Dependence of spin-dependent constants (in MeV) on ${\cal M}_{5/2}$.
The two branches are:
6355.4 MeV $< {\cal M}_{5/2} <$ 6382.5 MeV (left panel), and
6379.9 MeV $< {\cal M}_{5/2} <$ 6406.9 MeV (right panel).
Upper solid (black) ellipse: $a_1$; dashed single-valued line: $a_2$;
lower solid (red) ellipse: $b$; dashed (red) ellipse: $c$.  $\chi^2$ (defined
in the text): dash-dotted lines, nonzero only outside range of $M_{5/2}$
giving solutions.  Ellipses coalesce into a single line for values of $M_{5/2}$
giving minimum nonzero $\chi^2 \equiv \sum_{i=1}^5 (M_{i,{\rm exp}} -
M_{i,{\rm fit}})^2$.  Units of $\chi^2$ are in MeV$^2$.  The favored range of
$M_{5/2}$ is in the left-hand figure.
\label{fig:cts}}
\end{figure}

Varying ${\cal M}_{5/2}$ above 6350 MeV, we find solutions for the ranges 6355.4 MeV
$< {\cal M}_{5/2} <$ 6382.5 MeV and 6379.9 MeV $< {\cal M}_{5/2} <$ 6406.9 MeV, as shown in
the left-hand and right-hand panels of Fig.\ \ref{fig:cts}, respectively, with
two branches for $a_1,b,$ and $c$, and a single branch for $a_2$.


The constants in the figure may be compared with those favored in a fit to
excited $\Omega_c$ states \cite{Karliner:2017kfm}:
\beq \label{eqn:pred}
a_1 = 26.95 {\rm~MeV}~,~~
a_2 = 25.74 {\rm~MeV}~,~~
 b  = 13.52 {\rm~MeV}~,~~
 c  =  4.07 {\rm~MeV}~.
\eeq

It was argued in Ref.\ \cite{Karliner:2017kfm} that the hyperfine term $c$
should be no larger for the $\Omega_b$ system than for the excited $\Omega_c$
states.  In that case one selects the lower branch of the dashed (red) ellipse
in the left-hand panel, which is correlated with the upper branch
of the solid (black) upper ellipse, and favors values of $M_{5/2}$ in the
range of 6356 to 6366 MeV, with the most probable (lowest) value of $c$
for $M_{5/2} = 6358$ MeV.
Specifically, for  that value we find
\beq \label{eqn:linfit}
a_1 = 10.34 {\rm~MeV}~,~~
a_2 = 6.37 {\rm~MeV}~,~~
 b  = 7.02 {\rm~MeV}~,~~
 c  =  3.07 {\rm~MeV}~.
\eeq
The smaller value of $a_1$ in comparison with the value for $\Omega_c$ in
Eq.\ (\ref{eqn:pred}) shows the limitation of the extrapolation in Ref.\
\cite{Karliner:2017kfm}, which predicted them to be equal, but other
parameters are within their predicted ranges.

\section{\boldmath
Energy cost of a $P$-wave excitation \unboldmath
\label{sec:SP}}

We estimated the $P$-wave excitation energy for a $ss$ diquark bound to a $b$
quark with relative orbital angular momentum $L=1$ in Section V of Ref.\
\cite{Karliner:2017kfm}.  A crude value of 300 MeV was obtained.
It was necessary to anticipate the hyperfine splitting in the $S$-wave $bss$
ground state, as only the $\Omega_b(1/2^+)$ has been seen, with mass
$M(\Omega_b) = (6046.1 \pm1.7)$ MeV.  With an estimated hyperfine splitting
between $\Omega_b(1/2^+)$ and $\Omega^*_b(3/2^+)$ of 24 MeV, the spin-weighted
average of the $1/2^+$ and $3/2^+$ $S$-wave $bss$ states was estimated to be 6062
MeV.  Subsequently we noted \cite{Karliner:2018bms} that $P$-wave excitation
energies obeyed an approximate linear relation
\beq
\Delta E_{P-S} = \Sigma B + 417.4~{\rm MeV} - 0.214 \mu_R~,
\eeq
where $\Sigma B$ is the binding energy of the $(ss)$ 
diquark and $b$ quark, estimated to be
83.6 MeV, and $\mu_R$ is the reduced mass of the $ss$--$b$ system:
\beq
m_{ss} = 1098.8~{\rm MeV}~,~~m_b = 5041.8~{\rm MeV}~,~~\mu_R=902.2~{\rm MeV}~.
\eeq
With these inputs one finds $\Delta E_{P-S} = 308$ MeV, implying $\barM =6370$
MeV.  As we see above and below, under various assumptions the LHCb data imply
values a bit lower than this.

\section{Evaluation of predictions for \boldmath
$\Omega_b = \bss$ states \unboldmath\
\label{sec:omb}}

In addition to the predictions of Ref.\ \cite{Karliner:2017kfm} for the
hyperfine parameter $c$ and the S-P splitting, we predicted other parameters
for the $\Omega_b = bss$ states based on rescaling the fitted $\Omega_c$
values quoted in Eq.\ (\ref{eqn:pred}).  

\begin{itemize}

\item[(a)] The parameter $a_1$ was to be kept as in the $css$ system,
as it expresses the coefficient of ${\bold L} \cdot {\bold S_{(ss)}}$:
$a_1[\bss] = a_1[\css] = 26.95$ MeV.  On the other hand, for $M_{5/2}$
around the favored value of 6358 MeV, $a_1 \simeq 10$ MeV.  

\item[(b)] The spin-orbit parameter $a_2$ was expected to scale as the inverse
of the heavy quark mass:  $a_2[\bss] = (1708.8/5041.8)(25.74) = 8.72$ MeV,
where we have taken the charm and bottom quark masses from Ref.\
\cite{Karliner:2015ema}.  Its value in the present fit for $M_{5/2} \simeq
6358$ MeV is about 6 MeV (see the left-hand panel of Fig.\ \ref{fig:cts}).

\item[(c)] The tensor force parameter $b$ is found to be about 7 MeV,
well within the range of $\pm 20$ MeV anticipated in  \cite{Karliner:2015ema}.

\end{itemize}

\section{\label{sec:alt}Alternative interpretations}

Predictions for the negative-parity $\Omega_b$ states were made by
several authors \cite{Ebert:2011kk,Maltman:1980er,Ebert:2007nw,Roberts:2007ni,%
Garcilazo:2007eh,Migura:2006ep,Valcarce:2008dr,Yamaguchi:2014era,Bali:2015lka,%
Yoshida:2015tia,Shah:2016nxi,Wang:2017goq,Zhao:2017fov,Agaev:2017jyt}, in
papers prior to discovery of the excited $\Omega_c$ states, and by authors
commenting on those states \cite{Padmanath:2017, Wang:2017vnc,Wang:2017zjw,%
Chen:2017gnu,Aliev:2017led,Chen:2017sci,Wang:2017hej,Cheng:2017ove,
Agaev:2017lip,Santopinto:2018ljf,Ortiz-Pacheco:2020hmj}, 
including interpretations based on pentaquarks
\cite{Yang:2017rpg,Huang:2017dwn,Kim:2017jpx}.  Since the discovery of the
four narrow $\Omega_b$ states, several interpretations of them have been
proposed \cite{Chen:2020mpy,Liang:2020hbo,Liang:2020dxr,Wang:2020pri,%
Xiao:2020oif,Mutuk:2020rzm}.  These differ from one
another in their $J^P$ assignments and predicted widths. In Table \ref{tab:JP}
the first four columns denote states
quoted in the $(J,j$) basis (1/2,0), (1/2,1), (3/2,1), (3/2,1), (5/2,2), where
$j$ is the total angular momentum (spin and orbital angular momentum $L$) of
the $ss$ diquark, while the last two columns refer to states quoted in the
basis $^{2S+1}P_J = ^2P_{1/2}$, $^2P_{3/2}$, $^4P_{1/2}$, $^4P_{3/2}$,
$^4P_{5/2}$.  Typical errors in predictions are 10 to 20 MeV, except for the
QCD sum rule calculation \cite{Wang:2020pri}, whose errors are of order 100 MeV.
One should pay more attention to splitting among levels than their absolute
values.

A general pattern emerges from these calculations.  In the $j-j$ coupling
scheme, the single state with $j=0$ is deemed to be very wide 
(see Table \ref{tab:widths}), and hence
not observable in the current data set of LHCb \cite{Aaij:2020cex}.  The two
states with $j=1$ and the two with $j=2$ are expected to be narrow, for the
most part within the experimental resolution.  This behavior is not seen in
the case of the $\Omega_c$ states, where candidates for all five involving the
spin-one $ss$ pair in its ground state are seen \cite{Aaij:2017nav}.

Even if one assumes the reason for seeing four rather than five excited
$\Omega_b$ states is that the one with $j=0$ is very broad, there is no
unanimity on the order of the observed states.  That is the question we 
address in considering the effects of spin-orbit, tensor force, and spin-spin
couplings.  We have found a consistent solution in which it is the state with
$J^P=5/2^-$ that is missing in the data.  

We now repeat the exercise in which
the states are described by $j-j$ coupling and it is the one with $j=0$ whose
mass we vary in order to determine the parameters $a_1,a_2,b,c$.

\begin{table}
\caption{Comparison of predicted $J^P$ assignments of the narrow states
decaying to $\Xi_b^0 K^-$ reported by LHCb \cite{Aaij:2020cex}.  Masses in MeV.
\label{tab:JP}}
\begin{center}
\begin{tabular}{c c c c c c c} \hline \hline
  $J^P$ & \cite{Chen:2020mpy} & \cite{Liang:2020hbo}(a) & \cite{Wang:2020pri} &
 \cite{Xiao:2020oif} & \cite{Mutuk:2020rzm} & \cite{Ortiz-Pacheco:2020hmj}(b)\\
Coupling & $j-j$  &  $j-j$  &   $j-j$  &   $j-j$   &  $L-S$ & $L-S$ \\ \hline
$1/2^-$ & 6340(c) & 6339(c) &  6330(d) &  6316(c,d)&  6314  & 6305 \\ 
$1/2^-$ &  6340  &   6330   &   (c)?   &   6316(d) &  6330  & 6317 \\
$3/2^-$ &  6340  &   6340   &  6316(d) &   6340(d) &  6339  & 6313 \\
$3/2^-$ &  6350  &   6331   &  6350(d) &   6330(d) &  6342  & 6325 \\
$5/2^-$ &  6360  &   6334   &  6340(d) &   6350(d) &  6352  & 6338 \\
\hline \hline
\end{tabular}
\end{center}
\leftline{(a) Mass predictions taken from Ref.\ \cite{Ebert:2011kk}.}
\leftline{(b) Mass predictions taken from Ref.\ \cite{Santopinto:2018ljf}.}
\leftline{(c) State with $j=0$ predicted to be very broad and not seen by
LHCb.}
\leftline{(d) Experimental masses; proposed $J^P$ assignments.}
\end{table}

%
%
\begin{table}
\caption{Comparison of predicted widths (in MeV) of the narrow states
decaying to $\Xi_b \bar K$ in an S wave reported by LHCb \cite{Aaij:2020cex}.
\label{tab:widths}}
\begin{center}
\begin{tabular}{c c c c c c c} \hline \hline
  $J^P$ & \cite{Chen:2020mpy} & \cite{Liang:2020hbo} & \cite{Wang:2020pri} &
 \cite{Xiao:2020oif} & \cite{Mutuk:2020rzm} & \cite{Santopinto:2018ljf} \\
Coupling & $j-j$  &  $j-j$  &   $j-j$  &  $j-j$ & $L-S$ & $L-S$ \\ \hline
$1/2^-$ & $>1000$ & 871 (a) &   (b)    &   126  & 0.78  &  0.50 \\ 
$1/2^-$ &    0    &  --     &   (b)    &   --   & 3.18  &  1.14 \\
$3/2^-$ &    0    &  --     &   (b)    &   --   & 1.74  &  2.79 \\
$3/2^-$ &   0(c)) & 1.35(d) &   (b)    &   2.2  & 0.58  &  0.62 \\
$5/2^-$ &    0    & 2.98(e) &   (b)    &   3.4  & 2.83  &  4.28 \\
\hline \hline
\end{tabular}
\end{center}

\leftline{(a) For $\Omega_b(6316)$.  (1057,1146,1224) MeV for $\Omega_b%
(6330,6340,6350)$.}
\leftline{(b) Observed states at (6316,6330,6340,6350) MeV assigned $J^P =
(3/2^-,1/2^-,5/2^-,3/2^-)$}
\leftline{with comparable pole strengths in QCD sum rules.}
\leftline{(c) Plus predicted $4.7^{+6.1}_{-2.9}$ MeV for decay to $\Xi_b \bar
 K$ in a D wave.}
\leftline{(d) For $\Omega_b(6340)$.  (0.35,1.08,2.98) MeV for $\Omega_b%
(6316,6330,6350)$.}
\leftline{(e) For $\Omega_b(6350)$.  (0.35,1.08,1.85) MeV for $\Omega_b%
(6316,6330,6340)$.}
\end{table}

In order to determine mass splittings in the linearized $j-j$ coupling basis,
we use lowest-order perturbation theory in the inverse of $m_b$
\cite{Karliner:2017kfm}.  (If we kept nondiagonal terms in this basis the
outcome would be the same as in the $L-S$ coupling basis considered earlier.)
\bea
\Delta M(J=\frac12,j=0) &=& -2a_1~, \label{eqn:10}\\
\Delta M(J=\frac12,j=1) &=& \kern0.5em-a_1 -\frac12 a_2 -b -\frac12 c~,
\label{eqn:11} \\
\Delta M(J=\frac32,j=1) &=& \kern0.5em-a_1 +\frac14 a_2 + \frac12 b + \frac14 c~,
\label{eqn:31} \\
\Delta M(J=\frac32,j=2)&=&\phantom{-2}a_1-\frac34 a_2+\frac{3}{10} b-\frac34 c~,
\label{eqn:32} \\
\Delta M(J=\frac52,j=2) &=&\phantom{-2}a_1 +\frac12 a_2 - \frac15 b + \frac12 c~.
\label{eqn:52}
\eea
The independent parameters are $a_1$, $a_2+c$, and $b$, so the five mass
splittings obey two sum rules:
\begin{align}
2 \Delta M(1/2,1) + 4 \Delta M(3/2,1) &= \phantom{-}3 \Delta M(1/2,0)\,,
\label{eqn:lin1}\\
4 \Delta M(3/2,2) + 6 \Delta M(5/2,2) &= -5 \Delta M(1/2,0)\,.
\label{eqn:lin2}
\end{align}
where the first number refers to $J$ and the second to $j$.  We are assuming
that the unseen state is the one with mass $M(J=1/2,j=0)$.  Eliminating
$\Delta M(1/2,0)$ from the above two equations and recalling that each
$\Delta M \equiv M - \barM$, we find an expression for $\barM$ in terms of
the four observed masses, whose value depends on the permutation of the masses
 assigned to each $J^P$ level:
\beq
\barM = \frac16 M(1/2,1) + \frac13 M(3/2,1) + \frac15 M(3/2,2)
 + \frac{3}{10} M(5/2,2)~.
\eeq
We can now obtain $\Delta M(1/2,0)$ from Eq.\ (\ref{eqn:lin1}), and find
\beq
M(1/2,0) =  \frac13 \left[ 2M(1/2,1) + 4M(3/2,1) \right] - \barM~.
\eeq
The spin-dependent coefficients are
\bea
a_1 & = & - \frac12 \Delta M(1/2,0)~, \\
a_2 + c & = & \frac13 \left[ 3 \Delta M(1/2,0) -\Delta M(1/2,1)
 + 5 \Delta M(5/2,2) \right] ~,\\
b & = & \frac59 \left[ 3 \Delta M(3/2, 1) + \Delta M(3/2,2)
 - \Delta M(1/2, 0) \right] ~.
\eea
Table \ref{tab:perm} lists all 24 permutations of the masses of the observed
four levels with $(J,j) = (1/2,1),~(3/2,1),~(3/2,2),~(5/2,2)$, obtaining
parameters for each permutation.  We denote the order in which the observed
masses are monotonically increasing by the permutation 1 2 3 4. We then compare
each set with values estimated in Ref.\ \cite{Karliner:2017kfm} by extrapolation
from the excited $\Omega_c$ spectrum.  Some notable features are the following:

\begin{itemize}

\item The values of $a_1$ are half or less that estimated by extrapolating from
charm to bottom.  It probably pays to choose the largest possible (positive)
$a_1$.

\item The value of $a_2$ was estimated in \cite{Karliner:2017kfm} to be 8.72
MeV, while $c$ was estimated to be small, less than a few MeV.

\item Ref.\ \cite{Karliner:2017kfm} considered values of $b$ lying within
the range $-20 < b < 20$ MeV, satisfied by most sets in Table \ref{tab:perm}.

\item The value of $\barM$ varies within a narrow range around 6335 MeV, to be
compared with the crude estimate of 6362 MeV in Ref.\ \cite{Karliner:2017kfm},
 the central value of 6344 MeV obtained in Sec.\ III, and the value of 6370 MeV
found in Sec.\ \ref{sec:SP}.
 
\end{itemize}

With these considerations the set labeled by the permutation 1234 seems the
most satisfactory.  The observed levels at 6316, 6330, 6340, and 6350 MeV then
would correspond to the states with $(J,j) = (1/2,1),~(3/2,1),~(3/2,2),~
(5/2,2)$. respectively.

\section{Conclusions \label{sec:concl}}

We have interpreted the four narrow peaks seen by LHCb in the $\Xi_c^0 K^-$
mass distribution \cite{Aaij:2020cex} as $P$-wave excitations of a spin-1 $ss$
diquark with respect to a spin-1/2 $b$ quark.  While such a system is expected
to have five states --- two of spin 1/2, two of spin 3/2, and one of spin
5/2 ---, we advance arguments in favor of the spin-5/2 state being missed.
When the four observed levels are assigned $J^P = 1/2^-,1/2^-,3/2^-,3/2^-$
in order of ascending mass, solutions for spin-dependent parameters are
obtained for 6355.4 MeV $< M_{5/2} <$ 6382.5 MeV and 6379.9 MeV $< M_{5/2} <$
6406.9 MeV, with the lowest $\sim 10$ MeV of this range favored by consideration
of the derived spin-dependent parameters.

\begin{table}[H]
\caption{Parameters in MeV for all permutations of $J^P$ assignments of excited
$\Omega_b$ levels at 6316, 6330, 6339, and 6349 MeV.
\label{tab:perm}}
\begin{center}
\begin{tabular}{c r r r r r r} \hline \hline
\vrule width 0pt height 2.5ex
Permu-  &$\Delta M\kern0.5em\strut$&$M(1/2,0)\kern-0.7em\strut$&$\barM\kern1em\strut$& $a_1\kern1em\strut$ &
$a_2+c$ & $b$\kern1em\strut \\
tation\kern0.5em\strut  & (1/2,0) &        &        &          &         &     \\ \hline
1 2 3 4 &  -20.4  & 6315.2 & 6335.6 &  10.200  &  10.037 &    4.754 \\
1 2 4 3 &  -18.6  & 6316.2 & 6334.6 &   9.183  &  -3.523 &   11.534 \\
1 3 2 4 &  -10.4  & 6326.5 & 6336.9 &   5.181  &  18.402 &    6.845 \\
1 3 4 2 &   -6.4  & 6328.5 & 6334.9 &   3.223  &  -7.705 &   19.899 \\
1 4 2 3 &    2.5  & 6339.7 & 6337.2 &  -1.260  &  13.882 &   15.885 \\
1 4 3 2 &    4.4  & 6340.7 & 6336.3 &  -2.201  &   1.335 &   22.159 \\
2 1 3 4 &  -25.3  & 6307.9 & 6333.2 &  12.643  &   3.522 &  -11.535 \\
2 1 4 3 &  -23.3  & 6308.9 & 6332.2 &  11.626  & -10.038 &   -4.755 \\
2 3 1 4 &    0.4  & 6336.8 & 6336.4 &  -0.194  &  24.917 &   -6.186 \\
2 3 4 1 &    7.2  & 6340.2 & 6333.0 &  -3.618  & -20.736 &   16.641 \\
2 4 1 3 &   13.3  & 6350.0 & 6336.7 &  -6.635  &  20.397 &    2.854 \\
2 4 3 1 &   18.1  & 6352.4 & 6334.3 &  -9.042  & -11.696 &   18.901 \\
3 1 2 4 &  -18.4  & 6314.5 & 6332.9 &   9.193  &   7.704 &  -19.899 \\
3 1 4 2 &  -14.5  & 6316.4 & 6330.9 &   7.235  & -18.403 &   -6.846 \\
3 2 1 4 &   -2.7  & 6332.1 & 6334.8 &   1.374  &  20.735 &  -16.641 \\
3 2 4 1 &    4.1  & 6335.5 & 6331.4 &  -2.050  & -24.918 &    6.185 \\
3 4 1 2 &   22.1  & 6357.5 & 6335.5 & -11.027  &  12.033 &    0.763 \\
3 4 2 1 &   25.0  & 6359.0 & 6334.0 & -12.493  &  -7.514 &   10.536 \\
4 1 2 3 &   -8.9  & 6322.6 & 6331.5 &   4.447  &  -1.336 &  -22.159 \\
4 1 3 2 &   -7.0  & 6323.5 & 6330.6 &   3.506  & -13.883 &  -15.886 \\
4 2 1 3 &    6.7  & 6340.3 & 6333.5 &  -3.372  &  11.695 &  -18.901 \\
4 2 3 1 &   11.6  & 6342.6 & 6331.0 &  -5.779  & -20.398 &   -2.855 \\
4 3 1 2 &   18.7  & 6352.4 & 6333.8 &  -9.332  &   7.513 &  -10.537 \\
4 3 2 1 &   21.6  & 6353.9 & 6332.3 & -10.798  & -12.034 &   -0.764 \\
\hline \hline
\end{tabular}
\end{center}
\end{table}

An alternative explanation of the missing state offered by several authors
\cite{Chen:2020mpy,Liang:2020hbo,Wang:2020pri,Xiao:2020oif} envisions the
states as approximately diagonal in the $(J,j)$ basis, where $J$ is the total
angular momentum and $j$ is the $ss$-diquark's total (spin plus $L$) angular
momentum.  In this basis the $(1/2,0)$ state is predicted to be very wide and
hence not seen by LHCb.  In this case the most plausible set of spin-dependent
parameters is obtained when the four observed levels are assigned $(J,j) =
(1/2,1);~(3/2,1);~(3/2,2);~(5/2,2)$ in order of ascending mass.  Angular
distributions of decay products should be able to distinguish between this
scenario and the (favored) one in the preceding paragraph.

{\it Notes added:}  The molecular picture of excited $\Omega_b$ states
offered in Ref.\ \cite{Liang:2020dxr} proposes four states at 6405, 6427,
6465, and 6508 MeV, which they associate with (non-significant) enhancements
in the LHCb spectrum \cite{Aaij:2020cex} at 6402, 6427, 6468, and 6495 MeV.
A QCD sum rule calculation \cite{Agaev:2017ywp} finds four excited states:
(1P,1/2$^-$) at $6336 \pm 183$ MeV; (2S,1/2$^+$) at $6487 \pm 187$ MeV;
(1P,3/2$^-$) at $6301 \pm 193$ MeV; and (2S,3/2$^+$) at $6422 \pm 198$ MeV.
An unpublished undergraduate thesis \cite{Giustino2020} employs a simplified
quark model to predict $M(\Omega_b(5/2,2))$ in the range of 6364 to 6372 MeV.
A very recent analysis in the same spirit as ours predicts $M(1P,5/2^-)
\simeq 6352$ MeV \cite{Jia:2020vek}.

\section*{Acknowledgements}

We thank K. Azizi, H. Mutuk, M. Giustino, E. Oset, A. Parkhomenko,
and E. Santopinto for helpful communications.  The research of M.K. was
supported in part by NSFC-ISF grant No.\ 3423/19.

\end{document}